\documentclass[a4paper]{article}
\usepackage{authblk} % For multiple authors with different affiliations

\usepackage[english]{babel}
\usepackage[utf8x]{inputenc}
\usepackage[T1]{fontenc}

\usepackage[a4paper,top=3cm,bottom=2cm,left=3cm,right=3cm,marginparwidth=1.75cm]{geometry}

\usepackage{amsmath}
\usepackage{verbatim}
\usepackage{graphicx}
\usepackage{diagbox}
\usepackage{enumitem}
\usepackage{wrapfig}
\usepackage{multirow}
\usepackage{indentfirst}
\usepackage[colorlinks=true, allcolors=blue]{hyperref}

\title{Overview of TREC 2024 Biomedical Generative Retrieval (BioGen) Track
}

\author[1]{Deepak Gupta}
\author[1]{Dina Demner-Fushman}
\author[2]{William Hersh}
\author[2]{Steven Bedrick}
\author[3]{Kirk Roberts}
\affil[1]{National Library of Medicine, NIH}
\affil[2]{Oregon Health \& Science University}
\affil[3]{University of Texas Houston}

\date{}
\begin{document}
\maketitle
\section{Overview} \label{sec:into}
With the advancement of large language models (LLMs), the biomedical domain has seen significant progress and improvement in multiple tasks such as biomedical question answering, lay language summarization of the biomedical literature, clinical note summarization, etc. However, hallucinations or confabulations remain one of the key challenges when using LLMs in the biomedical and other domains. Inaccuracies may be particularly harmful in high-risk situations, such as medical question answering \cite{zhao2024heterogeneous}, making clinical decisions, or appraising biomedical research. Studies on evaluation of the LLMs’ abilities to ground generated statements in verifiable sources have shown that models perform significantly worse on lay-user generated questions \cite{wu2024well}, and often fail to reference relevant sources\cite{basaragin-etal-2024-know}. This can be problematic when those seeking information want evidence from studies to back up the claims from LLMs\cite{jamiaocae014}. Unsupported statements are a major barrier to using LLMs in any applications that may affect health. Methods for grounding generated statements in reliable sources along with practical evaluation approaches are needed to overcome this barrier.  Towards this, in our pilot task organized at TREC 2024, we introduced the task of reference attribution as a means to mitigate the generation of false statements by LLMs answering biomedical questions. 

In TREC 2024, we received a total of $30$ runs from $5$ teams. The majority of the runs utilized a two-stage retrieval augmented generation (RAG) approach to generate the answers with references. In the first stage, a retriever was used for relevant literature, and in the second stage, the LLMs were used to generate the answers with appropriate references.

\section{Task} \label{sec:task}
\noindent
\textbf{Reference Attribution:} Given a biomedical topic (question) and a stable version of PubMed documents, the task was to generate answers using Large Language Models (LLMs) or any other approaches. Each sentence\footnote{assertions/statements were approximated as sentences in the 2024 task} must be supported by up to three attributions (cited references), but no more than $30$ documents per answer. Each document should be referenced in the answers as PMIDs in square brackets as shown in Figure \ref{fig:topic}. The document sentences that support the answer statements are shown to illustrate the manual evaluation process.\\

\begin{figure}[!t]
\caption{Sample reference answer}
\fbox{\begin{minipage}{0.9\textwidth}
\noindent \textbf{Topic:} \textit{iron and ferritin levels in COVID-19} \\
\textbf{Question:} \textit{Why is transferrin and iron low in covid patients but ferritin high?} \\
\noindent \textbf{Narrative}: \textit{The patient is interested in the link between iron and infection, the role iron plays in infection and the implications for the COVID-19 course.} \\
\noindent \textbf{Sample Answer [adapted to patient-level health literacy]}: \textit{During infections, a battle for iron takes place between the human body and the invading viruses} [34389110]. \textit{The immune system cells need iron to defend the body against the infection [34389110]. The virus needs iron to reproduce} [31585922]. \textit{If iron balance is disrupted by the infection, ferritin levels are high} [34883281], \textit{which signals the disease is severe and may have unfavorable outcomes} [34048587, 32681497]. \textit{Ferritin is maintaining the body’s iron level }[18835072]. \textit{Some researchers believe that high levels of ferritin not only show the body struggles with infection, but that it might add to the severity of disease} [34924800]. \textit{To help covid patients, the doctors may lower the ferritin levels that are too high using drugs that capture iron} [32681497]. \\
\noindent \textbf{References [to be returned by the system] and supporting statements manually extracted from the documents during the evaluation}: 
%{32681497, 33380357, 34048587, 34389110, 34883281, 34924800, 34960751, 35008695, 35136706, 35240553} 

\begin{itemize}

\item 34389110 
A S1: During infections, a battle for iron takes place between the human host and the invading pathogens.
A S2: Once primed by the contact with antigen presenting cells, lymphocytes need iron to sustain the metabolic burst required for mounting an effective cellular and humoral response. 

\item 31585922 
A S3: Viruses depend on iron in order to efficiently replicate within living host cells.

\item 34883281
A S4: Ferritin was initially described to accompany various acute infections, both viral and bacterial, indicating an acute response to inflammation. 

\item 34048587
A S4: Elevated serum ferritin and IL-6 levels associated with increased mortality and with reduced mortality at ferritin levels <100 ng mL-1. 

\item 32681497
A S4: Numerous studies have demonstrated the immunomodulatory effects of ferritin and its association with mortality and sustained inflammatory process.

\item 18835072 
A S5: Ferritin, a major iron storage protein, is essential to iron homeostasis and is involved in a wide range of physiologic and pathologic processes.

\item 34924800
A S6: The inflammation cascade and poor prognosis of COVID-19 may be attributed to high ferritin levels.

\item 32681497
A S7: Iron chelation represents a pillar in the treatment of iron overload. In addition, it was proven to have an anti-viral and anti-fibrotic activity. Herein, we analyse the pathogenic role of ferritin and iron during SARS-CoV-2 infection and propose iron depletion therapy as a novel therapeutic approach in the COVID-19 pandemic.
\end{itemize}

\end{minipage}}
 \label{fig:topic}
\end{figure}
\section{Topics}

The first $40$ BioGen 2024 topics were developed using information requests submitted by self-identified non-clinicians to the National Library of Medicine. Additionally, $25$ topics were developed based on the collection\footnote{\url{https://github.com/kevinwu23/SourceCheckup}} that contains questions based on a documents from Mayo Clinic\footnote{\url{
https://www.mayoclinic.org/diseases-conditions}}, UpToDate\footnote{\url{https://www.uptodate.com/contents/table-of-contents/patient-education}}, and Reddit r/AskDocs\footnote{\url{http://www.reddit.com/r/askdocs}}. Example topics are shown in Table \ref{tab:example-topic}.

% Please add the following required packages to your document preamble:
% \usepackage{graphicx}
\begin{table}[]
\resizebox{\columnwidth}{!}{%
\begin{tabular}{l|l}
\hline
\textbf{Topic} & \textit{natural treatments for sleep apnea} \\ 
\textbf{Question} & \textit{Are there ways to prevent sleep apnea or treat it naturally?} \\ 
\textbf{Narrative} &\textit{ The patient is looking for natural remedies to prevent and treat sleep apnea} \\ \hline
 &  \\ \hline
\textbf{Topic} & \textit{Drug treatment for COPD stage 4} \\ 
\textbf{Question} & \textit{What drug or combination of drugs is most popular for treating stage 4 copd?} \\ 
\textbf{Narrative} & \begin{tabular}[c]{@{}l@{}}\textit{This patient is looking for the most effective and popular medications for }\\ \textit{advanced Chronic obstructive pulmonary disease (COPD)}.\end{tabular} \\ \hline
 &  \\ \hline
\textbf{Topic} & \textit{fbn1 mutation} \\ 
\textbf{Question} & \textit{what is fbn1 mutation?} \\ 
\textbf{Narrative} & \begin{tabular}[c]{@{}l@{}}\textit{A young woman who is planning to start a family has heard that one of her}\\  \textit{cousins had a mutation in the gene for fibrillin-1 (FBN1). She would like} \\ \textit{to know more about these mutations and the health problems that might} \\ \textit{be caused by the mutation}\end{tabular} \\ \hline
\end{tabular}%
}
\caption{Example topics from the TREC 2024 BioGen track.}
\label{tab:example-topic}
\end{table}

\section{Data}
The BioGen task used the latest annual baseline snapshot of Medline/PubMed, which goes approximately through the end of 2023. We provided a pre-processed set of $20,727,695$ PMIDs representing the abstracts in the 2023 snapshot. The participants were asked to cite and use the PMIDs while generating the answers available in this collection.
\section{Participating Teams and Methods}
\subsection{Participating Teams} We used the NIST-provided Evalbase platform\footnote{\url{https://ir.nist.gov/evalbase}} to release the
datasets, registration, and submissions of the participating teams. In total, $5$ teams participated in the BioGen track and submitted $30$ individual runs for the task.
\subsection{Methods} We summarize the approaches used by the participants as follows:
\begin{itemize}
    \item \textbf{ur-iw}: The team used multiple LLMs (\texttt{gemini-1.5-flash-001}, \texttt{ gpt4o-mini} etc.) for query expansion and utilized Elasticsearch to retrieve the relevant documents. After that, they extracted and reranked the snippets based on query relevance. The retrieved snippets were used to generate the answer using \texttt{gemini-1.5-flash-001} and \texttt{ gpt4o-mini} models.
     \item \textbf{ii\_research}: The team adopted BM25-based retrieval to generate a silver-standard dataset containing the query and relevant documents. The generated dataset was used to train a T5-based seq2seq ranking model. The trained model was used to rank the relevant documents against a question that was used to generate the answer using the \texttt{GPT-4o} model.
     \item \textbf{h2oloo}: In the first stage, the team used BM25 with Rocchio query expansion to rank the top 1000 documents against a query. After that, they used a multi-stage reranker to retrieve the top 20 relevant documents/snippets that were used to generate an answer using \texttt{Llama3.1} and \texttt{GPT-4o} models.
      \item \textbf{webis}: The team used retrieval followed by reranking to retrieve relevant documents; thereafter \texttt{GPT-4o} was used to generate the answers.
     \item \textbf{ielab}: The team followed a three-step approach where the relevant documents were retrieved and re-ranked based on the query and then LLMs were used to generate the answer. In the final stage, LLMs attributed each of the generated sentences to the list of retrieved documents. 
\end{itemize}
   
\section{Assessment}
\begin{figure}[h]
\caption{Evaluation steps}
\centering
 \label{fig:f1}
\includegraphics[width=0.9\textwidth]{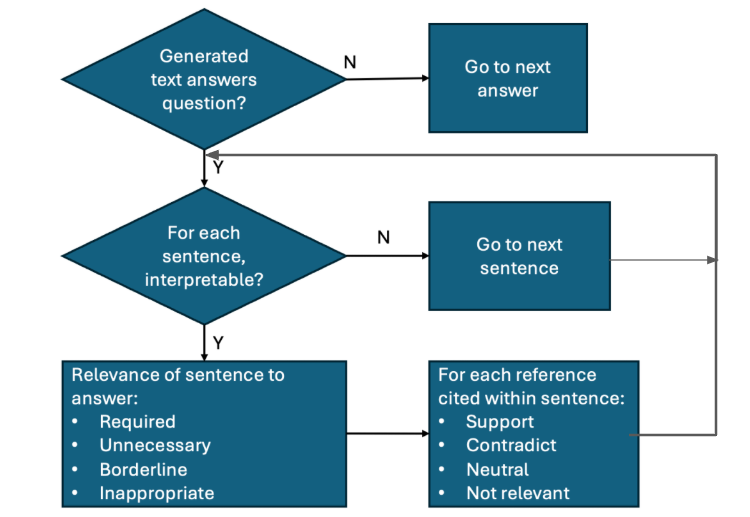}

\end{figure}

For the BioGen task, we focused on \textbf{(a)} reference attribution and \textbf{(b)} the quality and factuality of the text generated by LLMs to answer clinical questions asked by clinicians to \textbf{ (i)} satisfy their own information needs or \textbf{(ii)} answer health-related questions asked by their patients. For patients, we envisioned that the answers would be reviewed by clinicians and subsequently explained in plain language.
This evaluation aimed to verify the references attributed and the quality of LLM-generated answers. For the former, we evaluated how well the answers to clinical questions are supported by evidence provided by the models in the form of references. Here are the details of the two-part evaluation for the submissions (also shown in Figure~\ref{fig:f1}):
\begin{itemize}
    \item \textbf{Part 1: Evaluating Answer Alignment with Questions and Answer Quality and Completeness: } In the first step of the evaluation, we evaluated whether the generated text, taken as a whole, directly answered the question.  We then evaluated the relevance of the assertions in the answer sentences to the question. Each assertion in the generated answer was labeled with one of the following four labels: 
\begin{itemize}
    \item \textbf{Required}: The assertion ‘XXX’ is necessary to have in the generated answer for completeness of the answers.
    \item \textbf{Unnecessary}: The assertion ‘XXX’ is not required to have included in the generated answer. An assertion may be unnecessary for several reasons: it provides general information on the topic; it comments on the lack of information in provided documents; it recommends to see a doctor, while the task stated the patient has already contacted the provider, or the provider is asking the question.
    \item \textbf{Borderline}: If an assertion is relevant, possibly even “good to know,” but not required, the assertion may be marked borderline. For example, if the question is about most commonly used treatments, information about treatments in the early stages of clinical trials is not necessary. 
    \item \textbf{Inappropriate}: The assertion may harm the patient, e.g., if according to the answer, physical therapy reduces the pain level, but the patient experiences more pain due to hip mobilization, the patient may start doubting they are receiving adequate treatment.
\end{itemize}

    \item \textbf{Part 2: Evaluating Answer Alignment with Evidence Support:}
    
    In the second step, for each generated answer sentence, we assessed the referenced document(s) to determine each document's relation to the generated assertion, if any. We labeled each cited document with one of the four possible relations between the answer sentence and the document: `\textit{Supports}', `\textit{Contradicts}', `\textit{Neutral}', and `\textit{Not Relevant}'.

\begin{itemize}
    \item \textbf{Supports}: There is at least one sentence in the referenced document that supports/agrees with the statement, e.g.: \textit{``opioids were the mainstay of perioperative pain control''}. In addition, no other sentence in the document contradicts the statement. 

 \item \textbf{Contradicts}: There is at least one sentence in the referenced document that disagrees with the assertion or states its opposite, e.g.: \textit{``Increasing pain levels after the first week postoperatively, for 3 days, are most likely to be caused by the change to more extensive mobilization and physiotherapy in the rehabilitation unit."} (The answer in this case stated that the pain decreases steadily after the surgery.) 

 \item \textbf{Neutral}: The referenced document is topically relevant, but lacks any information to validate or invalidate the assertion.

 \item \textbf{Not relevant}: The referenced document is not relevant to the sentence.
\end{itemize}
\end{itemize}

\section{Evaluation}
We carried out an evaluation of approaches at several levels and along several axes.
\subsection{Answer quality}

\begin{itemize}
     \item \textbf{Answer Accuracy} (measured at the run level) -- measures how many of the answers to the total of 65 questions were deemed acceptable (judged as answering the question at least partially) for each run. %The accuracy is expected to be 90\% and above. 
      \begin{equation}
            \text{Accuracy}_{run} = \frac{\text{Number of Acceptable Answers}}{\text{Total Number of Questions (Topics)}}
        \end{equation}
    \item \textbf{Answer Completeness (Recall)} (answer level) -- measures how many of the answer aspects (pooled across all submitted runs from all participating teams) are covered in one answer to a question. For the initial evaluation, we group/cluster the sentences using the embeddings from SentenceTransformer\footnote{sentence-transformers/all-mpnet-base-v2} and SimCSE\footnote{princeton-nlp/sup-simcse-roberta-large} models. In a strict evaluation, only sentences that were judged required and supported by evidence will be considered. For a lenient evaluation, all sentences judged required will be considered. For a relaxed evaluation, the borderline sentences will be considered in addition to the required sentences. The number of aspects for the automated grouping is set to 10 using K-means clustering. 
    
        \begin{equation}
            \text{Completeness}_{answer} = \frac{\text{Number of Distinct Clusters Containing Sentences from Answer A}}{\text{Number of Clusters}}
        \end{equation}

        \item \textbf{Answer Precision} (answer level) -- measures how many of the assertions in the answer were judged required or acceptable. The strict, lenient, and relaxed evaluations also apply. 

          \begin{equation}
            \text{Precision}_{answer} = \frac{\text{Number of Generated Required Sentences}}{\text{Total Number of Generated Sentences}}
        \end{equation}

\item \textbf{Redundancy Score} (answer level): penalizes a system for generating \textit{unnecessary} answer sentences. It measures the informativeness of the generated answers. 
        \begin{equation}
            \text{Redundancy Score} =  \frac{\text{Number of Generated Unnecessary Sentences}}{\text{Total Number of Generated Sentences}}
        \end{equation}
 %        This score reflects the proportion of generated \textit{unnecessary} answer sentences among all generated answer sentences. 

\item \textbf{Irrelevant Score} (answer level): penalizes a system for generating \textit{inappropriate}/\textit{potentially harmful} answer sentences. %It measures how misleading or harmful the generated answers are. 
        \begin{equation}
            \text{Harmfulness Score} =  \frac{\text{Number of Generated Harmful Sentences}}{\text{Total Number of Generated Sentences}}
        \end{equation}
      %   This score reflects the proportion of generated \textit{inappropriate} answer sentences among all generated answer sentences. 
    \end{itemize} 
%    \end{enumerate}

\subsection{Citation Quality}
Answer statements may be supported or contradicted by the documents provided as references. Sentences may not include any references or include references that are only topically relevant or not relevant. The following metric capture the quality of references:

\begin{itemize}
  \item \textbf{Citation Coverage}: measures how well the required and borderline generated answer sentences are backed by the appropriate (judged as supports) citations. 
      \begin{equation}
      \footnotesize
          \text{Citation Coverage} =  \frac{\text{Number of Systems Generated Answer Sentences with One or More Supportive Citation}}{\text{Total Number of Generated Answer Sentences}}
       \end{equation}
         \item \textbf{Citation Support Rate}: assesses how well the system-predicted citations are aligned with the human-judged support citations.
    \begin{equation}
            \text{Citation Support Rate} =  \frac{\text{Number of \textit{Supports} Citations}}{\text{Total Number of Citations}}
        \end{equation}
          \item \textbf{Citation Contradict Rate}: penalizes the answers that are providing documents assessed as \textit{Contradicting} the statement. Note that in a fact-verification task, this measure may show how well a system is finding contradictory evidence. 
     \begin{equation}
            \text{Citation Contradict Rate} =  \frac{\text{Number of \textit{Contradict} Citations}}{\text{Total Number of Citations}}
       \end{equation}

\end{itemize}
\subsection{Document relevancy}
Pooling all documents judged relevant for a given topic, we can compute standard recall and precision. Note that relevant documents include documents judged as supporting, contradicting or being neutral. 

\begin{itemize}
\item Recall:
    \begin{equation}
            \text{Recall} = \frac{\text{Number of relevant retrieved documents}}{\text{all relevant documents}}
        \end{equation}

        \item Precision:
    \begin{equation}
            \text{Precision} = \frac{\text{Number of relevant retrieved documents}}{\text{Number of references provided}}
        \end{equation}

   \end{itemize}
   We computed each aforementioned metric for each question and then averaged over all the questions in the test collection to obtain the final scores.
% \bibliographystyle{unsrt}
% \bibliography{sample}

% Please add the following required packages to your document preamble:
% \usepackage{graphicx}
\begin{table}[]
\resizebox{\columnwidth}{!}{%
\begin{tabular}{l|l|c|c}
\hline
\multicolumn{1}{c|}{\textbf{Team Name}} & \multicolumn{1}{c|}{\textbf{Run Name}} & \begin{tabular}[c]{@{}l@{}}\textbf{Acceptable}\\ \textbf{Answers}\end{tabular} & \multicolumn{1}{c}{\textbf{Accuracy}} \\ \hline
\multirow{6}{*}{h2oloo} & listgalore\_gpt-4o\_ragnarokv5biogen\_top20 & 65 & 100 \\ 
 & listgalore\_l31-70b\_ragnarokv5biogen\_top20 & 65 & 100 \\ 
 & rl31-70b\_l31-70b\_ragnarokv5biogen\_top20 & 65 & 100 \\
 & rl31-70b\_gpt-4o\_ragnarokv5biogen\_top20 & 65 & 100 \\ 
 & rl31-70b\_gpt-4o\_ragnarokv5biogennc\_top20 & 65 & 100 \\ 
 & listgalore\_gpt-4o\_arenuggetsallyouneed.json & 65 & 100 \\ \hline
\multirow{10}{*}{ielab} & norarr.llm\_only\_2.llama3-70b\_fixed & 63 & 96.92 \\ 
 & rarr\_attronly.llm\_only\_2.llama3-70b\_fixed & 61 & 93.85 \\ 
 & rarr\_qgenfix.llm\_only\_2.llama3-70b\_fixed & 62 & 95.38 \\ 
 & rarr\_attrfix.llm\_only\_2.llama3-70b\_fixed & 64 & 98.46 \\ 
 & rarr\_qgen.llm\_only\_2.llama3-70b\_fixed & 62 & 95.38 \\ 
 & rarr\_attrfix\_custprompt.llm\_only\_3\_v2.llama3-70b\_fixed & 63 & 96.92 \\ 
 & rarr\_qgenfix\_custprompt.llm\_only\_3\_v2.llama3-70b\_fixed & 64 & 98.46 \\ 
 & rarr\_qgenfix.llm\_only\_2.llama3-8b\_fixed & 62 & 95.38 \\ 
 & norarr.llm\_only\_2.boolena.llama3-70b\_fixed & 58 & 89.23 \\ 
 & norarr.llm\_only\_2.llama3-8b\_fixed & 64 & 98.46 \\ \hline
ii\_research & iiresearch\_trec\_bio2024\_t5base\_run & 65 & 100 \\ \hline
\multirow{6}{*}{ur-iw} & ten-shot-gpt4o-mini & 65 & 100 \\ 
 & ten-shot-gemini-flash & 62 & 95.38 \\ 
 & ten-shot-gpt4o-mini-wiki & 61 & 93.85 \\ 
 & ten-shot-gemini-flash-wiki & 61 & 93.85 \\ 
 & zero-shot-gpt4o-mini & 64 & 98.46 \\ 
 & zero-shot-gemini-flash & 56 & 86.15 \\ \hline
\multirow{7}{*}{webis} & webis-1 & 53 & 81.54 \\ 
 & webis-2 & 43 & 66.15 \\
 & webis-3 & 39 & 60 \\ 
 & webis-5 & 43 & 66.15 \\ 
 & webis-gpt-4 & 54 & 83.08 \\ 
 & webis-gpt-6 & 60 & 92.31 \\ 
 & webis-gpt-1 & 65 & 100 \\ \hline
\end{tabular}%
}
\caption{Performance comparison of the submitted runs for the answer quality in terms of \textit{Accuracy} metric.}
\label{tab:res-acc}
\end{table}

% Please add the following required packages to your document preamble:
% \usepackage{graphicx}
\begin{table}[]
\resizebox{\columnwidth}{!}{%
\begin{tabular}{l|l|c|c|c}
\hline
\textbf{Team Name} & \textbf{Run Name} & \textbf{Precision} & \textbf{Redundancy} & \textbf{Harmfulness} \\ \hline
\multirow{6}{*}{h2oloo} & listgalore\_gpt-4o\_ragnarokv5biogen\_top20 & 80.5 & 14.07 & 0 \\ 
 & listgalore\_l31-70b\_ragnarokv5biogen\_top20 & 83.53 & 12.8 & 0.38 \\ 
 & rl31-70b\_l31-70b\_ragnarokv5biogen\_top20 & 82.08 & 14.37 & 0.26 \\ 
 & rl31-70b\_gpt-4o\_ragnarokv5biogen\_top20 & 82.87 & 12.32 & 0 \\ 
 & rl31-70b\_gpt-4o\_ragnarokv5biogennc\_top20 & 80.9 & 16.31 & 0 \\ 
 & listgalore\_gpt-4o\_arenuggetsallyouneed.json & 82.15 & 12.97 & 0.93 \\ \hline
\multirow{10}{*}{ielab} & norarr.llm\_only\_2.llama3-70b\_fixed & 89 & 4.03 & 0 \\ 
 & rarr\_attronly.llm\_only\_2.llama3-70b\_fixed & 84.31 & 3.51 & 0 \\ 
& rarr\_qgenfix.llm\_only\_2.llama3-70b\_fixed & 83.28 & 8.74 & 0.38 \\ 
 & rarr\_attrfix.llm\_only\_2.llama3-70b\_fixed & 87.31 & 6.95 & 0.82 \\ 
 & rarr\_qgen.llm\_only\_2.llama3-70b\_fixed & 87.51 & 5.18 & 0 \\ 
 & rarr\_attrfix\_custprompt.llm\_only\_3\_v2.llama3-70b\_fixed & 87.54 & 4.49 & 0.77 \\ 
 & rarr\_qgenfix\_custprompt.llm\_only\_3\_v2.llama3-70b\_fixed & 90.54 & 4.28 & 0 \\ 
 & rarr\_qgenfix.llm\_only\_2.llama3-8b\_fixed & 84.7 & 6.6 & 1 \\ \
 & norarr.llm\_only\_2.boolena.llama3-70b\_fixed & 80.28 & 5.21 & 0.38 \\ 
& norarr.llm\_only\_2.llama3-8b\_fixed & 90.68 & 4.99 & 0.31 \\ \hline
ii\_research & iiresearch\_trec\_bio2024\_t5base\_run & 69.44 & 23.41 & 1.54 \\ \hline
\multirow{6}{*}{ur-iw} & ten-shot-gpt4o-mini & 83.93 & 12.9 & 0 \\ 
 & ten-shot-gemini-flash & 75.88 & 15.21 & 0.48 \\ 
 & ten-shot-gpt4o-mini-wiki & 75.62 & 15.37 & 0 \\ 
 & ten-shot-gemini-flash-wiki & 74.07 & 17.13 & 0 \\ 
 & zero-shot-gpt4o-mini & 81.98 & 12.59 & 0.22 \\ 
 & zero-shot-gemini-flash & 68.4 & 13.09 & 0 \\ \hline
\multirow{7}{*}{webis} & webis-1 & 65.54 & 11.9 & 0 \\ 
 & webis-2 & 52.82 & 12.05 & 0 \\ 
 & webis-3 & 52.79 & 6.51 & 0 \\ 
 & webis-5 & 55.59 & 8.64 & 0 \\ 
 & webis-gpt-4 & 64.23 & 16.54 & 0 \\ 
 & webis-gpt-6 & 62.56 & 26.41 & 0 \\ 
 & webis-gpt-1 & 81.67 & 14.74 & 0 \\ \hline
\end{tabular}%
}
\caption{Evaluation of submitted runs for answer quality, focusing on \textit{Precision}, \textit{Redundancy}, and \textit{Harmfulness} metrics.}
\label{tab:res-quality}
\end{table}

% Please add the following required packages to your document preamble:
% \usepackage{graphicx}
\begin{table}[]
\resizebox{\columnwidth}{!}{%
\begin{tabular}{l|l|c|c|c}
\hline
\textbf{Team Name} &\textbf{ Run Name} & \textbf{Recall (S+R)} &\textbf{ Recall (R)} & \textbf{Recall (R+B)} \\ \hline
\multirow{6}{*}{h2oloo} & listgalore\_gpt-4o\_ragnarokv5biogen\_top20 & 34.31 & 38.15 & 40.15 \\ 
 & listgalore\_l31-70b\_ragnarokv5biogen\_top20 & 34 & 38.46 & 41.08 \\ 
 & rl31-70b\_l31-70b\_ragnarokv5biogen\_top20 & 34.15 & 38.15 & 38 \\ 
 & rl31-70b\_gpt-4o\_ragnarokv5biogen\_top20 & 32.92 & 39.38 & 41.69 \\ 
 & rl31-70b\_gpt-4o\_ragnarokv5biogennc\_top20 & 0 & 40.31 & 41.23 \\ 
 & listgalore\_gpt-4o\_arenuggetsallyouneed.json & 43.08 & 46.31 & 46.15 \\ \hline
\multirow{10}{*}{ielab} & norarr.llm\_only\_2.llama3-70b\_fixed & 26 & 26.62 & 27.69 \\ 
 & rarr\_attrfix\_custprompt.llm\_only\_3\_v2.llama3-70b\_fixed & 25.23 & 26 & 27.69 \\
 & rarr\_qgenfix\_custprompt.llm\_only\_3\_v2.llama3-70b\_fixed & 23.38 & 26.31 & 27.38 \\ 
 & rarr\_attronly.llm\_only\_2.llama3-70b\_fixed & 24.15 & 24.62 & 27.23 \\ 
 & rarr\_attrfix.llm\_only\_2.llama3-70b\_fixed & 25.69 & 26 & 27.69 \\ 
 & rarr\_qgenfix.llm\_only\_2.llama3-8b\_fixed & 23.08 & 26.31 & 27.23 \\ 
 & rarr\_qgenfix.llm\_only\_2.llama3-70b\_fixed &23.38&	26	&26.46 \\
 & rarr\_qgen.llm\_only\_2.llama3-70b\_fixed & 22.46 & 24 & 25.08 \\ 
 & norarr.llm\_only\_2.boolena.llama3-70b\_fixed & 21.69 & 22.77 & 23.54 \\ 
 & norarr.llm\_only\_2.llama3-8b\_fixed & 24.31 & 27.08 & 27.69 \\ \hline
ii\_research & iiresearch\_trec\_bio2024\_t5base\_run & 1.38 & 25.08 & 26.15 \\ \hline
\multirow{6}{*}{ur-iw} & ten-shot-gpt4o-mini & 28.77 & 38.15 & 39.54 \\ 
 & ten-shot-gemini-flash & 27.54 & 33.54 & 32.31 \\ 
 & ten-shot-gemini-flash-wiki & 24.77 & 32.77 & 32.92 \\ 
& zero-shot-gpt4o-mini & 26.77 & 37.85 & 39.54 \\ 
 & zero-shot-gemini-flash & 25.08 & 32.15 & 33.54 \\ 
& ten-shot-gpt4o-mini-wiki & 25.08 & 34.92 & 35.85 \\ \hline
\multirow{7}{*}{webis} & webis-1 & 7.38 & 9.69 & 11.08 \\ 
 & webis-5 & 3.23 & 8.15 & 8.62 \\ 
 & webis-gpt-6 & 7.08 & 16.31 & 17.08 \\ 
webis & webis-gpt-1 & 0 & 20.15 & 20.15 \\ 
 & webis-2 & 4.77 & 7.69 & 8.31 \\ 
 & webis-3 & 5.85 & 7.85 & 8.31 \\ 
 & webis-gpt-4 & 7.08 & 14 & 14.77 \\ \hline
\end{tabular}%
}
\caption{Comparison of submitted runs based on answer quality using the Recall metric. The abbreviations are as follows: (\textbf{S+R})-- only answer sentences that were judged required and supported by evidence were considered to cluster, (\textbf{R})-- answer sentences that were judged required were considered to cluster, (\textbf{R+B})--  answer sentences that were judged either required or borderline were considered to cluster.
}
\label{tab:res-recall}
\end{table}

% Please add the following required packages to your document preamble:
% \usepackage{graphicx}
\begin{table}[]
\resizebox{\columnwidth}{!}{%
\begin{tabular}{l|l|c|c|c}
\hline
\textbf{Team Name} & \textbf{Run Name }& \begin{tabular}[c]{@{}l@{}}\textbf{Citation}\\ \textbf{Coverage}\end{tabular}  & \begin{tabular}[c]{@{}c@{}}\textbf{Citation Support}\\ \textbf{Rate}\end{tabular} & \begin{tabular}[c]{@{}c@{}}\textbf{Citation Contradict}\\ \textbf{Rate}\end{tabular}\\ \hline
\multirow{6}{*}{h2oloo} & listgalore\_gpt-4o\_ragnarokv5biogen\_top20 & 82.79 & 79.97 & 1.76 \\ 
 & listgalore\_l31-70b\_ragnarokv5biogen\_top20 & 78.47 & 77.02 & 2.43 \\ 
 & rl31-70b\_l31-70b\_ragnarokv5biogen\_top20 & 81.14 & 77.41 & 2.88 \\ 
 & rl31-70b\_gpt-4o\_ragnarokv5biogen\_top20 & 80.32 & 78 & 2.49 \\ 
 & rl31-70b\_gpt-4o\_ragnarokv5biogennc\_top20 & 0 & 0 & 0 \\ 
 & listgalore\_gpt-4o\_arenuggetsallyouneed.json & 88.58 & 62.54 & 1.93 \\ \hline
\multirow{10}{*}{ielab} & norarr.llm\_only\_2.llama3-70b\_fixed & 92.21 & 73.62 & 1.89 \\ 
 & rarr\_attronly.llm\_only\_2.llama3-70b\_fixed & 86.54 & 70.84 & 4.69 \\ 
 & rarr\_qgenfix.llm\_only\_2.llama3-70b\_fixed & 85.77 & 65.01 & 1.93 \\ 
 & rarr\_attrfix.llm\_only\_2.llama3-70b\_fixed & 89.05 & 73.22 & 6.13 \\ 
 & rarr\_qgen.llm\_only\_2.llama3-70b\_fixed & 83.1 & 67.55 & 0.97 \\ 
 & rarr\_attrfix\_custprompt.llm\_only\_3\_v2.llama3-70b\_fixed & 88.56 & 71.98 & 3.53 \\ 
 & rarr\_qgenfix\_custprompt.llm\_only\_3\_v2.llama3-70b\_fixed & 87.54 & 67.45 & 3.16 \\ 
 & rarr\_qgenfix.llm\_only\_2.llama3-8b\_fixed & 81.74 & 68 & 2.12 \\ 
 & norarr.llm\_only\_2.boolena.llama3-70b\_fixed & 81.36 & 60 & 1.55 \\ 
 & norarr.llm\_only\_2.llama3-8b\_fixed & 88.18 & 55.16 & 2.45 \\ \hline
ii\_research & iiresearch\_trec\_bio2024\_t5base\_run & 5 & 11.54 & 0 \\ \hline
\multirow{7}{*}{ur-iw} & ten-shot-gpt4o-mini & 66.35 & 68.63 & 1.88 \\ 
 & ten-shot-gemini-flash & 72.94 & 64.02 & 0.26 \\ 
& ten-shot-gpt4o-mini-wiki & 60.61 & 65.77 & 0.19 \\ 
 & ten-shot-gemini-flash-wiki & 65.89 & 56.59 & 1.26 \\ 
 & zero-shot-gpt4o-mini & 64.13 & 60.56 & 1.55 \\ 
& zero-shot-gemini-flash & 63.06 & 58.13 & 1.35 \\ \hline
\multirow{7}{*}{webis} & webis-1 & 52.51 & 41.56 & 3.82 \\ 
 & webis-2 & 34.1 & 41.79 & 4.62 \\ 
 & webis-3 & 42.97 & 45 & 5.38 \\ 
 & webis-5 & 29.36 & 34.62 & 0.77 \\ 
 & webis-gpt-4 & 42.18 & 68.72 & 2.31 \\ 
 & webis-gpt-6 & 32.31 & 37.51 & 3.64 \\
 & webis-gpt-1 & 0 & 0 & 0 \\ \hline
\end{tabular}%
}
\caption{Comparison of submitted runs based on citation quality using multiple metrics.}
\label{tab:res-citation-quality}
\end{table}

% Please add the following required packages to your document preamble:
% \usepackage{graphicx}
\begin{table}[]
\resizebox{\columnwidth}{!}{%
\begin{tabular}{l|l|c|c}
\hline
\textbf{Team Name} &\textbf{ Run Name} & \textbf{Recall} & \textbf{Precision} \\ \hline
\multirow{6}{*}{h2oloo} & listgalore\_gpt-4o\_ragnarokv5biogen\_top20 & 12.91 & 88.03 \\ 
 & listgalore\_l31-70b\_ragnarokv5biogen\_top20 & 14.31 & 90.04 \\ 
 & rl31-70b\_l31-70b\_ragnarokv5biogen\_top20 & 14.23 & 89.21 \\ 
 & rl31-70b\_gpt-4o\_ragnarokv5biogen\_top20 & 12.63 & 88.64 \\ 
 & rl31-70b\_gpt-4o\_ragnarokv5biogennc\_top20 & 0 & 0 \\ 
 & listgalore\_gpt-4o\_arenuggetsallyouneed.json & 24.12 & 77.08 \\ \hline
\multirow{10}{*}{ielab} & norarr.llm\_only\_2.llama3-70b\_fixed & 9.36 & 83.64 \\ 
 & rarr\_attronly.llm\_only\_2.llama3-70b\_fixed & 10.37 & 81.75 \\ 
 & rarr\_qgenfix.llm\_only\_2.llama3-70b\_fixed & 9.16 & 76.93 \\ 
 & rarr\_attrfix.llm\_only\_2.llama3-70b\_fixed & 11.29 & 86.59 \\
 & rarr\_qgen.llm\_only\_2.llama3-70b\_fixed & 8.63 & 77.98 \\
 & rarr\_attrfix\_custprompt.llm\_only\_3\_v2.llama3-70b\_fixed & 10.56 & 81.34 \\ 
 & rarr\_qgenfix\_custprompt.llm\_only\_3\_v2.llama3-70b\_fixed & 9.26 & 76.47 \\ 
 & rarr\_qgenfix.llm\_only\_2.llama3-8b\_fixed & 7.25 & 80.79 \\ 
 & norarr.llm\_only\_2.boolena.llama3-70b\_fixed & 8.52 & 74.22 \\ 
 & norarr.llm\_only\_2.llama3-8b\_fixed & 9.01 & 72.52 \\ \hline
ii\_research & iiresearch\_trec\_bio2024\_t5base\_run & 0.6 & 22.82 \\ \hline
\multirow{6}{*}{ur-iw}& ten-shot-gpt4o-mini & 9.06 & 73.62 \\ 
 & ten-shot-gemini-flash & 8.89 & 69.23 \\ 
 & ten-shot-gpt4o-mini-wiki & 8.4 & 69.72 \\ 
& ten-shot-gemini-flash-wiki & 8.73 & 60.22 \\ 
& zero-shot-gpt4o-mini & 8.29 & 70.64 \\ 
 & zero-shot-gemini-flash & 6.51 & 61.22 \\ \hline
\multirow{7}{*}{webis} & webis-1 & 3.03 & 50.77 \\ 
 & webis-2 & 1.15 & 52.31 \\ 
 & webis-3 & 1.16 & 46.92 \\ 
 & webis-5 & 0.96 & 40.51 \\ 
 & webis-gpt-4 & 2.28 & 77.69 \\ 
 & webis-gpt-6 & 2.93 & 54.62 \\ 
 & webis-gpt-1 & 0 & 0 \\ \hline
\end{tabular}%
}
\caption{Evaluation of submitted runs for document relevancy focusing on \textit{Precision} and \textit{Recall} metrics.}
\label{tab:res-doc-relevance}
\end{table}

\section{Results and Discussion} We evaluated the performance of the submitted runs on multiple levels for answer quality (Tables \ref{tab:res-acc}, \ref{tab:res-quality}, \ref{tab:res-recall}), citation quality (Table \ref{tab:res-citation-quality}) and document relevance (Table \ref{tab:res-doc-relevance}). For the answer accuracy (deemed acceptable for a given question), we found most of the runs achieved more than 90\% accuracy. For precision of the answer quality, the run norarr.llm\_only\_2.llama3-8b\_fixed achieved a maximum score of $90.68\%$, while the run webis-2 (where the answer was generated by the \texttt{Mistral-7B-Instruct-v0.3} model) recorded the lowest precision of the $52.82$. For redundancy, we found the run rarr\_attronly.llm\_only\_2.llama3-70b\_fixed recorded the lowest redundancy score of $3.51$. The run used \texttt{Llama-3-70B-Instruct} model to generate the answer. The listgalore\_gpt-4o\_arenuggetsallyouneed run obtained the highest recall in multiple settings (S+R, R, R+B). The best citation coverage of $92.21\%$ was obtained by norarr.llm\_only\_2.llama3-70b\_fixed run, in which  \texttt{Llama-3-70B-Instruct} was asked to attribute each of the generated answer sentences explicitly. The best citation support rate (CSR) of $79.97\%$ was achived by listgalore\_gpt-4o\_ragnarokv5biogen\_top20. The best recall and precision scores of $24.12$ and $90.04$ for document relevancy were obtained by the listgalore\_gpt-4o\_arenuggetsallyouneed and listgalore\_l31-70b\_ragnarokv5biogen\_top20 runs. 

\section{Conclusion}
This overview of the TREC 2024 BioGen track discussed the tasks, datasets,
evaluation metrics, participating systems and their performance. We evaluated the performance of the submitted runs at multiple levels (answer, citation and documents) using the traditional metrics. Most of the teams used the two-step approach where they first retrieved the documents with lexical search method (BM25), followed by reranker to obtained the top-k relevant documents/snippets. In the second stage, LLMs were used to generate the answer citing appropriate documents. We hope that introducing the task has created ground truth datasets for fostering research toward designing systems that generate answers to health-related questions grounded with appropriate citations, which, in turn,  provide a trusted and reliable source to support the generated assertions in the answer. 

\section*{Acknowledgments}
 This research was supported in part by the Division of Intramural Research (DIR) of the National Library of Medicine (NLM), National Institutes of Health.  
 Bill Hersh, Kate Fultz Hollis, and Kirk Roberts were supported by grant R01LM011934 from the National Library of Medicine.
 The authors thank Kate Fultz Hollis, OHSU  and Srishti Kapur, Centaur Labs for managing the manual evaluation process at OHSU and Centaur labs, respectively.

\bibliographystyle{unsrt}
\bibliography{sample}

\end{document}